\documentclass{aa}
\usepackage{graphicx}

\begin{document}
\title{Magnetic helicity evolution during the solar activity cycle:
observations and dynamo theory}
\author{N.\, Kleeorin \inst{1}, K.\, Kuzanyan \inst{2}, D.\, Moss
\inst{3}, I.\, Rogachevskii \inst{1}, D.\, Sokoloff \inst{4} \and
H.\, Zhang \inst{5} } \offprints{I. Rogachevskii}
\institute{Department of Mechanical Engineering, Ben-Gurion
University of Negev, POB 653, 84105 Beer-Sheva, Israel\\
\email{nat@menix.bgu.ac.il; gary@bgumail.bgu.ac.il} \and IZMIRAN,
Troitsk,
Moscow Region 142190, Russia\\
\email {kuzanyan@dnttm.ru}\and Department of
Mathematics, University of Manchester, Manchester M13 9PL, UK \\
\email{moss@maths.man.ac.uk} \and Department of Physics, Moscow
State
University, Moscow 119992, Russia \\
\email{sokoloff@dds.srcc.msu.su} \and National Astronomical
Observatories, Chinese Academy of Sciences, Beijing 100012, China\\
\email{hzhang@bao.ac.cn} }
\date{Received ; accepted}
\authorrunning{Kleeorin et al.}

\titlerunning{Magnetic helicity in solar cycle}

\abstract{We study a simple model for the solar dynamo in the
framework of the Parker migratory dynamo, with a nonlinear dynamo
saturation mechanism based on magnetic helicity conservation
arguments. We find a parameter range in which the model
demonstrates a cyclic behaviour with properties similar to that of
Parker dynamo with the simplest form of algebraic
$\alpha$-quenching. We compare the nonlinear current helicity
evolution in this model with data for the current helicity
evolution obtained during 10 years of observations at the Huairou
Solar Station of China. On one hand, our simulated data
demonstrate behaviour comparable with the observed phenomenology,
provided that a suitable set of governing dynamo parameters is chosen.
On the other hand, the observational data are shown to be rich
enough to reject some other sets of governing parameters. We
conclude that, in spite of the very preliminary state of
the observations and the crude nature of the model, the idea of using
observational data to constrain our ideas concerning magnetic
field generation in the framework of the solar dynamo appears
promising.
}
\maketitle

\section{Introduction}
The solar activity cycle is widely believed to be connected with
dynamo action which occurs somewhere inside the solar convective
zone or even in the overshoot layer. Starting from the seminal paper of
Parker (1955), various dynamo models have been suggested for the
solar cycle (see e.g. R\"udiger \& Brandenburg 1995; Sofia et
al. 1998; Tavakol et al. 2002; Brooke et al. 2002; also Blackman \&
Brandenburg 2003, whose dynamo model of solar cycle also includes
magnetic helicity balance). These models exploit particular
parameterizations for sources of the dynamo activity, i.e. the
$\alpha$-effect, which in turn is connected with the mean hydrodynamic
helicity $\chi^v$ of the convective motions, and acts in conjunction with the
nonuniform rotation $\Omega$. If the dynamo action is strong enough,
a dynamo wave propagating somewhere inside the convective shell is
excited. It is necessary to include some saturation mechanism to get
a (quasi)stationary wave which can be compared with the observed
activity cycle, instead of a dynamo wave with an exponentially
growing amplitude. In principle, the phenomenology of the solar
cycle can be reproduced using a very primitive $\alpha$-quenching
model of dynamo saturation, with the energy of the dynamo generated
magnetic field achieving approximate equipartition with the kinetic
energy of the random motions.

A deeper treatment of solar dynamo saturation requires however
some ideas concerning the physical processes that give rise to
quenching of the generation mechanism. A scenario of dynamo
saturation which is now widely discussed is connected with the
concept of magnetic helicity. The point is that the weakest link in
the dynamo self-excitation chain, i.e. $\alpha$, is a pseudoscalar
quantity and cannot be directly connected with the magnetic energy,
which is a scalar (not pseudoscalar) quantity. A magnetic helicity
$\chi^m$ can however be introduced to describe the level of
magnetic field mirror-asymmetry and this quantity can be
associated with the magnetic part of $\alpha$, i.e. $\alpha^m$,
which is thought to be responsible for $\alpha$-quenching.

The magnetic helicity $\chi^m$ is an integral of motion for the
ideal MHD equation, similar to the hydrodynamic helicity $\chi^v$ which
is conserved in the hydrodynamical case. During the solar
activity cycle, magnetic helicity is redistributed between the
large and small scale magnetic field. Based on this concept, a
governing equation for $\alpha^m$ has been suggested (Kleeorin \&
Ruzmaikin 1982; Kleeorin \& Rogachevskii 1999). Together with the
mean-field dynamo equations, this equation has solutions in form
of a propagating steady dynamo wave (Kleeorin et al. 1994, 1995;
Covas et al. 1998; Blackman \& Brandenburg 2002).

For a long time, it was impossible to observe either the magnetic
helicity $\chi^m$ or the hydrodynamic helicity $\chi^v$ and these
values, crucial for dynamo theory, were taken from theoretical
estimates only. In last decade, basic progress here has been
achieved and the first observations of magnetic helicity in active
regions on the solar surface have been obtained (Pevtsov et al.
1994, 1995; Zhang \& Bao 1998, 1999; Canfield \& Pevtsov 1998;
Longcope et al. 1998). It is possible to some extent to isolate a
latitudinal distribution of magnetic helicity averaged over a
solar cycle (Zhang et al. 2002) as well as to follow the temporal
evolution of magnetic helicity averaged over latitude. The obvious
aim now is to confront predictions of dynamo theory concerning the
latitudinal distribution of magnetic helicity and its evolution
during a solar cycle with the corresponding observational data;
this is the aim of the present paper. When carrying out our
investigation, we take into account that the available data
concerning magnetic helicity of the solar magnetic field are still
quite uncertain, and it would be unrealistic to expect that more
or less fine details can be isolated using this data.
Correspondingly, we restrict ourself to a very crude theoretical
model, that we confront with observations. Specifically, we
simplify the mean-field dynamo equations at the level of the
Parker migratory dynamo equations, and include the algebraic
$\alpha$-quenching and the dynamic $\alpha$-quenching associated
with magnetic helicity evolution as the only saturation
mechanisms. Both hypotheses are obvious simplifications and there
is no problem in principle in including many more realistic features
into our dynamo model. However we consider that to be a topic for
further work.

The other point to be clarified from the very beginning is the
following. Magnetic helicity can be understood as a measure of the
linkage of magnetic lines and it is necessary to reconstruct the
complete 3D magnetic field structure to deduce this helicity from
observations.
Clearly, this is a very complicated observational problem and
various intermediate quantities such as current helicity $ \langle
\vec{b} \cdot (\vec{\nabla} {\bf \times} \vec{b}) \rangle $, i.e.
the linkage between electric current lines, are used to this end
(where $\vec{b}$ represents magnetic fluctuations). These quantities are
useful in theoretical studies of dynamo saturation also and we use
them below. Our work needs a clear distinction between such
concepts as $\alpha$-effect and the corresponding helicity, and between
helicities of total magnetic field, large-scale magnetic field and
small-scale magnetic field; these distinctions can be neglected to some extent
in other areas of dynamo theory.

\section{Magnetic and current helicity data obtained at the
Huairou Solar Observing Station}

The averaged value of the small-scale magnetic helicity, i.e.
$\langle \vec{a} \cdot \vec{b} \rangle$, evidently would be a convenient
quantity to confront with a dynamo saturation scenario based on a
magnetic helicity conservation argument. In practice however, the
vector potential $\vec{ a}$, being a non-gauge invariant quantity is
inconvenient observationally, and it is the current helicity $
\langle {\vec{ b}} \cdot ({\bf \nabla} \times {\vec{ b}}) \rangle =
\langle b_x ({\bf \nabla} \times {\vec{ b}})_x \rangle + \langle b_y
({\bf \nabla} \times {\vec{ b}})_y \rangle + \langle b_z ({\bf
\nabla} \times {\vec{ b}})_z \rangle $ which can be extracted from
the observations (here $x,y,z$ are local cartesian coordinates
connected with a point on the solar surface and the $z$-axis is
normal to the surface). The observations are restricted to active
regions on the solar surface and we obtain information concerning
the surface magnetic field and helicity only. Monitoring of solar
active regions while they are passing near to the central meridian of the
solar disc enables observers to determine the full magnetic field
vector. The observed magnetic field is subjected to further
analysis to obtain the value ${\bf \nabla} \times {\vec{ b}}$.
Because it is calculated from the surface magnetic field
distribution, the only electric current component that can be
calculated is $({\bf \nabla} \times {\vec{ b}})_z$. As a consequence of these
restrictions, the observable quantity is

\begin{equation}
H_c = \langle b_z ({\bf \nabla} \times {\vec{ b}})_z \rangle \; .
\label{observ}
\end{equation}

Because the surface magnetic field is almost force-free, it is
also useful to consider the magnetic field twist $a_{\rm ff}$ (Woltjer,
1958) which
is defined as the proportionality coefficient between magnetic field
and electric current (${\bf j} = a_{\rm ff} {\bf b}$). The observational
restrictions discussed above imply that the observational equivalent
of twist is

\begin{equation}
a_{\rm ff} = j_z/b_z.
\label{twist}
\end{equation}

The observational data used in our analysis were obtained at
the Huairou Solar Observing station of the National Astronomical
Observatories of China. The magnetographic instrument based on the
FeI 5324 \AA \, spectral line determines the magnetic field values
at the photospheric level. The data are obtained from a CCD
camera with $512 \times 512$ pixels over the whole magnetogram,
whose entire size is comparable with the size of an active region,
as well as with the depth of the solar convective zone (about
$2\times 10^{8}$\,m). However, because of the observational technique,
the line-of-sight field component $b_z$ can be determined with a
much higher precision than the transverse components ($b_x$ and
$b_y$). There are a number of other observational difficulties
such as in resolving the so-called `$180^\circ$ ambiguity' in
the direction of the transverse field etc. The observational technique
is described in detail by Wang et al. (1996),
see also Abramenko et al. (1996).

An observational programme to reveal the values of the twist and
the current helicity density over the solar surface requires a
systematic approach, both to the monitoring of magnetic fields in
active regions and to the data reduction, in order to reduce the
impact of noise. This work has been carried out by a number of
research groups (e.g. Seehafer 1990; Pevtsov et al. 1994; Rust \&
Kumar 1996; Abramenko et al. 1997; Bao \& Zhang 1998; Kuzanyan et
al. 2000). While this work is still in progress, the largest
systematic data-set of active regions presently available consists
of 422 active regions over the 10 years 1988-1996 (Bao \& Zhang
1998). We use averages and confidence intervals calculated from
these data (Table~1). Note that both the averaged quantities of
twist and current helicity density are positive/negative over
Southern/Northern solar hemispheres respectively, and thus obey
the so-called hemispheric rule (Table~2).

\begin{table}
\caption[]{ The first column gives the central latitude $\Theta =
90^\circ - \theta$ of the data bin, with the averaging interval in
brackets, the twist $\langle a_{\rm ff} \rangle$ is measured in
units of $10^{-8}\, {\rm m}^{-1}$, the current helicity $\langle
H_c \rangle$ in units of $10^{-3} {\rm G}^2 {\rm m}^{-1}$, and $N$
is the number of active regions involved in the analysis. The
errors correspond to the 95\% confidence level, i.e. about two
standard deviations. }
\begin{flushleft}
\begin{tabular}{|c|r|r|l|}
\hline $\Theta$ & $\langle a_{\rm ff} \rangle$ & $\langle H_c
\rangle$ &
$N$ \\
\hline
28 (24 - 32) & $-0.4\pm 1.2$ & $-1.6\pm 1.7$ & 18\\
20 (16 - 24) & $-0.9\pm0.8$ & $-0.9\pm0.4$ & 51\\
14 (12 - 16) & $-1.7\pm1.3$ & $-0.6\pm0.4$ & 34\\
10 (8 - 12) & $-2.2\pm0.6$ & $-0.4\pm0.2$ & 49\\
4 (0 - 8) & $-1.9\pm0.8$ & $-0.6\pm0.2$ & 44\\
$-4$ ($-8$ -0) & $0.3\pm0.7$ & $0.7\pm0.5$ & 31\\
$-10$ ($-12$ - $-8$) & $1.2\pm0.7$ & $0.7\pm0.4$ & 59\\
$-14$ ($-16$ - $-12$) & $0.9\pm0.7$ & $0.9\pm0.7$ & 46\\
$-20$ ($-24$ - $-16$) & $1.0\pm0.8$ & $0.4\pm0.2$ & 68\\
$-28$ ($-32$ - $-24$) & $1.6\pm1.7$ & $0.5\pm0.9$ & 14\\
\hline
\end{tabular}\end{flushleft}
\end{table}
\begin{table}
\caption[]{The data of Table~1 binned by hemisphere and year of
observation.}
\begin{flushleft}
\begin{tabular}{|c|c|c|r|}
\hline
$T$ & $\langle a_{\rm ff} \rangle$ & $\langle H_c \rangle$ & $N$ \\
\hline
\multicolumn{4}{|c|}{North}\\
\hline
1988-89 & $-1.1 \pm 0.8$ & $-1.0\pm 0.5$ & 50\\
1990-91 & $-1.0 \pm 0.7$ & $-1.0\pm 0.5$ & 61\\
1992-93 & $-2.1 \pm 0.7$ & $-0.7\pm 0.3$ & 45\\
1994-95 & $-2.6 \pm 0.9$ & $-0.3\pm 0.1$ & 34\\
1996-97 & $-1.2 \pm 1.0$ & $-0.2\pm 0.2$ & 9 \\
\hline
\multicolumn{4}{|c|}{South}\\
\hline
1988-89 & $1.0 \pm 1.2$ & $0.2\pm 0.3$ & 38\\
1990-91 & $0.9 \pm 0.7$ & $0.8\pm 0.6$ & 65\\
1992-93 & $1.2 \pm 0.5$ & $0.9\pm 0.3$ & 77\\
1994-95 & $0.7 \pm 0.9$ & $0.1\pm 0.1$ & 35\\
1996-97 & $0.3 \pm 2.0$ & $0.2\pm 0.3$ & 8 \\
\hline
\end{tabular}\end{flushleft}
\end{table}

Observations at Huairou Observing Station also give dopplergrams
of velocity fields ${\vec v}$ over active regions from the FeI
5324 \AA \, spectral line in the photosphere and from the
$H_\beta$ line in the chromosphere. The processing of these data
could provide values of $v_z b_z$ (which is related to the
so-called cross-helicity) in the foreseeable future. However, at
present this quantity is not available for statistical studies.

\section{The dynamo model}
We describe the solar dynamo by means of the mean-field equation
(e.g. Moffatt 1978; Parker 1979; Krause \& R\"adler 1980;
Zeldovich et al. 1983) which in general form is

\begin{equation}
\frac{\partial {\vec{B}}}{\partial t}= \vec{\nabla} {\bf \times}
(\vec{V} {\bf \times} \vec{B} + \vec{\cal E} - \eta_0 \,
\vec{\nabla} {\bf \times} \vec{B}) \;, \label{E1}
\end{equation}
where $ {\vec{ V}} $ is a mean velocity (e.g. the differential
rotation), $ \eta_0 $ is the magnetic diffusion due to the
electrical conductivity of the fluid, $ \vec{\cal E} = \langle
{\vec{ u}} \times {\vec{ b}} \rangle $ is the mean electromotive
force, $ {\vec{ u}} $ and $ {\vec{ b}} $ are fluctuations of the
velocity and magnetic field respectively, and angular brackets
denote averaging over an ensemble of fluctuations. The
electromotive force $\vec{\cal E}$ can be separated into several
contributions, which include the $\alpha$-effect, turbulent
magnetic diffusivity $\eta$ and other terms such as the magnetic
turbulent diamagnetic effect. For now, we restrict ourself to the
two first terms, and consider $\alpha$ and $\eta$ to be isotropic
quantities. We take the turbulent diffusivity as a prescribed
quantity and take into account the nonlinearity of the
$\alpha$-effect only, i.e. we use the parameterization

\begin{equation}
\vec{\cal E} = \alpha(\vec{B}) {\vec B} - \eta \vec{\nabla} {\bf
\times} \vec{B} \;, \label{E2}
\end{equation}
where $\alpha$ depends in principle on the entire evolution of the
magnetic field, rather on its value in a given instant. This
dependence is described by an evolution equation (such as Eqs.
(\ref{A3}) and (\ref{A6A}) below.)

Using spherical coordinates $r, \theta, \phi$, we represent an
axisymmetric mean magnetic field as $ \vec{B} = B_\phi
\vec{e}_{\phi} + \vec{\nabla} {\bf \times} (A \vec{e}_{\phi})$.
Following Parker (1955) we consider dynamo action in a thin
convective shell, average $A$ and $B_\phi$ over the depth of the
convective shell and consider these quantities as functions of
colatitude $\theta$ only. Then we neglect the curvature of the
convective shell
and replace it by a flat slab to get the following
equations (we drop the suffix on $B_\phi$ for the sake of brevity)

\begin{eqnarray}
{{\partial B} \over {\partial t}} &=& g D \sin \theta {{\partial
A} \over {\partial \theta}} + {{\partial ^2 B} \over {\partial
\theta^2}} - \mu ^2 B \;,
\label{eqB}\\
{{\partial A} \over {\partial t}} &=& \alpha B + {{\partial^2 A}
\over {\partial \theta^2}} - \mu^2 A \;, \label{eqA}
\end{eqnarray}
(see Appendix A). Here we measure lengths in units of the solar radius
and time in units of a diffusion time based on the solar radius and
turbulent magnetic diffusivity. The terms $-\mu^2 B$ and $-\mu^2
A$ represent the role of turbulent diffusive losses in the
radial direction -- the value $\mu=3$ corresponds to a convective
zone with a thickness of about 1/3 of the solar radius. $g =
\partial \Omega /
\partial r $ is the radial shear of differential rotation. We
neglect any latitudinal dependence of the rotation curve as well
as the link between poloidal and toroidal magnetic field via the
$\alpha$-effect, and so consider a simple $\alpha\omega$-dynamo.
$\alpha$ and $g$ are normalized with respect to their maximal
values and incorporated into the dimensionless dynamo number $D$,
which gives the intensity of the dynamo action (see Sokoloff et
al. 1995 for mathematical details of the derivation of Eqs.
(\ref{eqB}), (\ref{eqA}) from Eqs. (\ref{E1}), (\ref{E2})).

These equations are obviously oversimplified. Starting from the
fundamental paper of Parker (1955) they can be used however to
reproduce basic qualitative features of solar and stellar activity
and appear to be viable for this purpose. Taking into account
the nature of the approach, we use the simplest profiles of
dynamo generators compatible with symmetry requirements, i.e.
$\alpha(\theta) = \cos \theta$ and $g =1$.

The points $\theta = 0$ and $\theta = 180^\circ$ correspond to
North and South poles respectively. We take here zero boundary
conditions for $A$ and $B$. (Because we neglect the convective
shell curvature, these boundary conditions necessarily are
approximate.) In principle, we could use a slightly more elaborate
version of Eqs.~(\ref{eqB}), (\ref{eqA}) which take into account
some curvature effects, so the diffusion term becomes formally
singular at the poles and a more realistic finiteness condition
can be exploited (see Galitski \& Sokoloff 1999). Magnetic
helicity data are available for middle latitudes and the
equatorial region ($ -30^\circ <\Theta < 30^\circ$ where the
latitude $\Theta = 90^\circ - \theta $ and $\Theta = 0$
corresponds to the equator) only and so here we are not very
interested in details of dynamo wave behaviour near to the poles.
We keep a factor $\sin \theta$ in Eq. (\ref{eqB}) which reflects
the fact that the length of the parallels $\theta = {\rm const}$
vanishes at the poles (Kuzanyan \& Sokoloff 1995). Neglecting this
term results in an unphysical coupling between the dynamo wave
behaviour near to the pole and near to the equator.

We are interested in dynamo waves propagating from middle solar
latitudes towards the equator. This corresponds to negative dynamo
numbers provided $\alpha$ is chosen to be positive in the Northern
hemisphere and $g$ is positive near to the solar equator. According
to various models, the ranges of $|D| \approx 10^3 - 10^6$ can be
considered as realistic for the solar case.

We nondimensionalize the dynamo equations by measuring length in
units of the solar radius $R$, time in units of the turbulent
magnetic diffusion time $R^2 / \eta$, and the differential
rotation $\delta\Omega$ in units of the maximal value of $\Omega$.
$\alpha $ is measured in units of the maximum value of the
hydrodynamic part of the $ \alpha $-effect.

It is convenient to present the dynamo number as $D = R_\alpha
R_\omega ,$ where $R_{\alpha} = \alpha R / \eta \sim 1 - 200 $ and
$R_\omega = \delta \Omega R^2 / \eta \sim (1 - 4) \times 10^{3} $
(where typical values of parameters have been used for these
estimates) represent the  contributions of the $\alpha$-effect and
differential rotation, respectively. We use the equipartition
magnetic field $ B_\ast = u \sqrt{4 \pi \rho}$ as the unit of
magnetic field. The vector potential of the poloidal field $A$ is
measured in units of $ R_{\alpha} R B_\ast $, the density $ \rho $
normalized to its value at the bottom of the convective zone, and
the basic scale of the turbulent motions $l$ and turbulent
velocity $u$ at the scale $l $ are measured in units of their
maximum values through the convective region. The
magnetic Reynolds number $ \, {\rm Rm} = l u / \eta_0 $
is defined using these maximal values. $\eta = l u / 3 $
is an estimate for the turbulent diffusivity.

We stress that all physical ingredients of the model vary
more-or-less strongly with the depth $h$ below the solar surface
and we have to use some average quantities in the Parker dynamo
equations. We use mainly estimates of governing parameters taken
from models of the solar convective zone (see, e.g., Spruit 1974
and Baker \& Temesvary 1966; more modern treatments make little
difference to these estimates). In particular, at depth $ h \sim 2
\times 10^{10}$ cm, $ {\rm Rm} \sim 2 \cdot 10^9 ,$ $ \, u \sim 2
\times 10^3 $ cm s$^{-1}$, $ l \sim 8 \times 10^9$ cm, $ \rho \sim
2 \times 10^{-1}$ g cm$^{-3} ,$ $\, \eta \sim 5.3 \times 10^{12} $
cm$^2$s$^{-1}$. The density stratification scale is estimated here
as $ \Lambda_\rho = \rho / |\nabla \rho| \sim 6.5 \times 10^9$ cm
and the equipartition mean magnetic field $B_\ast = 3000 $ G. In
the upper part of the convective zone, say at depth $ h \sim 2
\times 10^7$ cm, these parameters are $ {\rm Rm} \sim 10^5 ,$ $\,
u \sim 9.4 \times 10^4 $ cm s$^{-1}$, $ l \sim 2.6 \times 10^7$
cm, $ \, \rho \sim 4.5 \times 10^{-7}$ g cm$^{-3} ,$ $\, \eta \sim
0.8 \times 10^{12} $ cm$^2$ s$^{-1}$ and $ \Lambda_\rho \sim 3.6
\times 10^7$ cm; the equipartition mean magnetic field is $B_\ast
= 220 $ G here. This estimate for the equipartition magnetic field
at the base of the convection zone $ (B_\ast = 3000 $ G) is
roughly consistent with the magnetic field strength in sunspots
(about 1 kG). However obviously it should be distinguished from
the mean magnetic field at the solar surface; a deeper discussion
of this distinction is outside of the scope of the paper. For the
Parker migratory dynamo, the toroidal magnetic field usually
dominates and below we ignore the poloidal magnetic field when
calculating the magnetic energy.

\section{The nonlinearities}
\label{nonlin}
A key idea of the dynamo saturation scenario exploited below is a
splitting of the total $\alpha$ effect into the hydrodynamic
($\alpha^v $) and magnetic ($\alpha^m $) parts

\begin{equation}
\alpha (r, \theta) = \alpha^v + \alpha^m \;, \label{AA3}
\end{equation}
as first suggested by Frisch et al. (1975). We need to
parameterize both contributions, $\alpha^v$ and $\alpha^m$, in
terms of the magnetic field components and helicities. Two types
of effect should be taken into account. First of all, the link
between $\alpha$-effect and the relevant helicities can be
modified by the dynamo-generated magnetic field. Correspondingly,
we introduce quenching functions $\phi_v$ ($\alpha^v = \chi^v
\phi_{v}$ with $\chi^v = - (\tau /3) \langle \vec{u}
\cdot(\vec{\nabla} {\bf \times} \vec{u}) \rangle$) and $\phi_m$
($\alpha^m = \chi^c \phi_{m},$ and $\chi^c$ is defined below) to
obtain $ \alpha = \chi^v \phi_v + \chi^c \phi_m ,$ where $ \tau $
is the correlation time of the turbulent velocity field. The
second problem to be addressed is that magnetic helicity is not a
very convenient quantity because it involves a gauge-noninvariant
quantity, i.e. the vector potential. We connect magnetic helicity
with the current helicity $\langle \vec{b} \cdot (\vec{\nabla}
{\bf \times} \vec{b}) \rangle$, by using the approximation of
locally homogeneous turbulent convection (see Kleeorin \&
Rogachevskii 1999). Then we need to obtain a quantity of suitable
dimension, and introduce the density $\rho$ to obtain the
correctly dimensioned $\chi^{c} \equiv (\tau / 12 \pi \rho)
\langle \vec{b} \cdot (\vec{\nabla} {\bf \times} \vec{b})
\rangle$. Thus,

\begin{equation}
\alpha (r, \theta) = \chi^v \phi_{v} + \chi^c \phi_{m} \; .
\label{A3}
\end{equation}

The quenching functions $\phi_{v}$ and $\phi_{m}$ in
Eq.~(\ref{A3}) are given by

\begin{eqnarray}
\phi_{v}(B) &=& (1/7) [4 \phi_{m}(B) + 3 L(B)] \;,
\label{A4} \\
\phi_{m}(B) &=& {3 \over {8B^2}} [1 - \arctan (\sqrt{8} B) /
\sqrt{8} B] \; \label{A5}
\end{eqnarray}
(see Rogachevskii \& Kleeorin 2000, 2001), where $ L(B) = 1 - 16
B^{2} + 128 B^{4} \ln (1 + 1/(8B^2)) .$ Thus $\phi_{v} = 1/(4B^2)$
and $\phi_{m} = 3/(8B^2)$ for $B \gg 1/3 ;$ and $\phi_{v} =
1-(48/5)B^2$ and $\phi_{m} = 1-(24/5)B^2$ for $B \ll 1/3 .$ Here
$ \chi^v $ and $ \chi^c$ are measured in units of the maximal
value of the $\alpha$-effect.

The function $\phi_{v}$ describes conventional quenching of the $
\alpha $ effect. A simple form of such a quenching, $\phi_v = 1/(1
+ B^{2}) ,$ was introduced long ago (see, e.g. Iroshnikov 1970).
This form is quite close to the more sophisticated form presented
in Eq.~(\ref{A4}). The magnetic part $\alpha^m$ includes two types
of nonlinearity: the algebraic quenching described by the function
$\phi_{m}$ (see e.g. Field et al. 1999; Rogachevskii \& Kleeorin
2000, 2001) and the dynamic nonlinearity which is determined by
Eq. (\ref{A6A}).

The quenching of the $\alpha$-effect is caused by the direct and
indirect modification of the electromotive force by the mean
magnetic field. The indirect modification of the electromotive
force is caused by the effect of the mean magnetic field on the
velocity fluctuations and on the magnetic fluctuations, while the
direct modification is due to the effect of the mean magnetic
field on the cross-helicity (see, e.g., Rogachevskii \& Kleeorin
2000, 2001).

We can calculate also the cross-helicity $\langle \vec{u} \cdot
\vec{b} \rangle$ which may in the future be compared with
observational data

\begin{eqnarray}
\langle \vec{u} \cdot \vec{b} \rangle = (\eta / 2) [3
\Lambda_u^{-1} B_r + \phi_{\rm ch}(B) \, (\vec{B} \cdot
\vec{\nabla}) B^2] \,,
\label{CH1}
\end{eqnarray}
where $\Lambda_u^{-1} = |\vec{\nabla} \langle \vec{u}^2 \rangle| /
\langle \vec{u}^2 \rangle ,$ $\, \phi_{\rm ch}(B) = (2/35 B^2)
[(15 + 224 B^2) \phi_{m}(2B) + 6 L(2B) - 21] ,$ and $\phi_{\rm
ch}(B) = - 128 / 5 $ for $ B \ll 1/3 ,$ and $\phi_{\rm ch}(B) = -
3 \pi / (20 \sqrt{2} \, B^3) $ for $ B \gg 1/3 .$ When
deriving Eq.~(\ref{CH1}) we used Eqs. (A14)-(A17) and (A21)
of Rogachevskii \& Kleeorin (2001).

Now we need to average Eq.~(\ref{A3}) over the depth of the
convective zone. The first term in the averaged equation seems to
be determined by the values taken at some sort of mean position in
the convective zone, while the situation concerning the second term is much
less clear, because the density used to calculate $\chi^c$
decreases strongly with radius. The clarification of this problem
is obviously beyond the Parker approximation; however to address
this problem as far as it possible here we introduce a
phenomenological parameter $\sigma$ by

\begin{equation}
\alpha (\theta) = \chi^v \phi_{v} + \sigma \chi^c \phi_{m} \;,
\label{E31}
\end{equation}
where the helicities and quenching functions are associated with
some sort of mean position in the convective zone. We emphasize that
below we consider $\sigma$ as a free parameter in the context of
the averaging process used to derive the Parker equations;
probably we can only safely assert that $\sigma\ga 1$.

For the sake of brevity of notation, we keep in Eq. (\ref{E31})
the same notation as in Eqs.~(\ref{AA3}) and~(\ref{A3}). We stress
that now we consider a parameterization for helicities which
depends on colatitude $\theta$ only. Of course, this is not more
than a phenomenological description, to be improved in more
detailed models of the nonlinear solar dynamo.

The function $\chi^c(\vec{B})$ is determined by a dynamical
equation which follows from the conservation law for magnetic
helicity (see Kleeorin \& Rogachevskii 1999). A general
dimensional form of this equation reads

\begin{equation}
{{\partial \chi^{c}} \over {\partial t}} + {\chi^c \over T} = -{1
\over 9 \pi \, \eta \, \rho_\ast} \, (\vec{\cal E} {\bf \cdot}
\vec{B} + \vec{\nabla} \cdot \vec{\Phi}) + \kappa \Delta \chi^c \;.
\label{A6A}
\end{equation}
Here $ \vec{\Phi} = C \chi^v \phi_{v} \vec{B}^2 \, l^2 \vec{e}_r\,
/ \Lambda_\rho $ is a nonadvective flux of the magnetic helicity
(here $\vec{e}_r$ is the unit vector in radial direction), $ -
\kappa \vec{\nabla} \chi^c $ is the diffusive flux of the magnetic
helicity (see Kleeorin \& Rogachevskii 1999; Kleeorin et al. 2000;
2002; 2003), and $ T = l^2 / \eta_0 $ is the relaxation time of
magnetic helicity. Magnetic helicity transport through the
boundary of a dynamo region is reported by Chae (2001) to be
observable at the solar surface. We also take into account that
for an axisymmetric problem the term which determines the
advective flux of the magnetic helicity, $ \vec{\nabla} \cdot
(\vec{V} \chi^{c}), $ vanishes ($ \vec{V} = \vec{e}_{\phi} \,
\Omega \, r \sin\theta$ is the differential rotation). The
parameter $C$ is a numerical coefficient and $\kappa$ is of order
$\eta$. In principle, these parameters can be calculated given
some model of convection, however we here take into account our
real level of knowledge and keep them as free parameters. Note
that in estimating the helicity flux $\vec{\Phi}$ we have to
include density gradients in the radial direction which are
neglected in other parts of the analysis.

The physical meaning of Eq.~(\ref{A6A}) is that the total magnetic
helicity is a conserved quantity and if the large-scale magnetic
helicity grows with magnetic field, the evolution of the small-scale
helicity should somehow
compensate this growth. Compensation mechanisms
include dissipation and various kinds of transport.

The dynamical equation~(\ref{A6A}) for the function
$\chi^c(\vec{B})$ in nondimensional form in the context of the
Parker migratory dynamo reads

\begin{eqnarray}
{{\partial \chi^c} \over {\partial t}} &+& (T^{-1} + \kappa
\mu^2){\chi^c} = \biggl({2 R \over l} \biggr)^2 \biggl( {{\partial
A} \over {\partial \theta}} {{\partial B} \over {\partial \theta}}
- \, B {{\partial^2 A} \over {\partial \theta ^2}}
\nonumber\\
&& - \alpha B^2 + 2 \mu^2 A B + C B^2 \phi_v \chi^v(\theta)
\biggl) + \kappa {{\partial ^2 \chi^c} \over {\partial \theta^2}}
\; \label{E9}
\end{eqnarray}
(see Appendix A), where we have averaged Eq.~(\ref{A6A}) over the depth
of the convective zone, so that the averaged value of $T^{-1}$ is

\begin{eqnarray}
T^{-1} &=& H^{-1} \int T^{-1}(r) \,d r \sim {\Lambda_l \, R^2 \,
\eta_0 \over H \, l^2 \, \eta} \approx 0.2 - 0.5 \;, \label{TE31}
\end{eqnarray}
$\Lambda_l$ is the characteristic scale of the variations $l ,$ $
\, T(r) = (\eta / R^{2}) (l^2 / \eta_0) $ is the nondimensional
relaxation time of magnetic helicity, and the quantities
$\Lambda_l , \, \eta_0 , \, l $ in Eq.~(\ref{TE31}) are associated
with the upper part of the convective zone. The parameters $C$ and
$ \kappa $ can be chosen as $C \sim (0.1 -    1) \times 10^{-1} $
and $ \kappa = 0.1 -    1 .$ The factor $ 10^{-1} $ in the
coefficient $C$ arises from the parameter $ (l^2 / \Lambda_\rho) /
R \sim 0.1 $ (see Appendix A).

\section{Results}
We simulated numerically the model of the nonlinear solar dynamo
based on the Parker approximation and conservation of magnetic
helicity arguments, as presented in previous sections. We found
that the model gives a stable nonlinear wave-type solution similar
to the solar cycle phenomenology from the generation threshold of
the nonlinear system ($D_{\rm nonl,crit} \approx -940$ with
$\sigma = \kappa =1, C=0.1$) up to $D=-10^4$. It is quite
interesting that a slightly stronger generation is needed to get
stable nonlinear oscillations with a nonvanishing amplitude than
to excite the linear dynamo, where $D_{\rm crit} \approx -910$.
The temporal behaviour of nonlinear dynamo waves is quite similar
to that with the simple algebraic $\alpha$-quenching, and we give
here, instead of quite standard plots, details of the cycle
parameters in Table~3. A feature of the activity cycles
illustrated by this Table is a transition from singly to doubly
periodic solutions with growth of dynamo intensity.

\begin{table}
\caption[]{ Parameters of activity cycles: $B_{\rm max}$ is the
value of the dimensionless amplitude of the toroidal magnetic
field in units of the equipartition field (estimated as 3000 G at
the bottom of the convective zone), $T_c$ is the dimensionless
cycle length, SP and DP denote singly and doubly periodic
solutions respectively; `runs away' means that no stable finite
amplitude solution was found, although in some cases there is a
long pseudo-stable initial phase. The other governing parameters
are $\sigma = 1$, $T = 3$, $(2R/l)^2 = 300$. For the DP solutions,
the amplitude of the stronger cycles and the shorter period are
given. }

\begin{flushleft}
\begin{tabular}{|l|l|r|l|l|l|}
\hline
$C$ & $\kappa$ & $D$ & $B_{\rm max}$ & $T_c$ & \\
\hline
0.1 & 1 & $-10^3$ & 0.12 & 0.37 & SP\\
0.1 & 1 & $-10^4$ & 5.28 & 0.25 & DP\\
0.1 & 1 & $-2\times 10^4$ & - & - & runs away\\
0.1 & 0.1 & $-10^3$ & 0.09 & 0.345 & SP\\
0.1 & 0.1 & $-10^4$ & - & - & runs away\\
0.1 & 3 & $-10^3$ & 0.13 & 0.355 & SP\\
0.1 & 3 & $-3\times 10^3$ & 1.47 & 0.46 & SP\\
0.1 & 3 & $-10^4$ & - & - & runs away\\
0.01 & 1 & $-10^3$ & 0.11 & 0.35& SP\\
0.01 & 1 & $-3\times 10^3$ & 1.35 & 0.37 & SP\\
0.01 & 1 & $-10^4$ & 4.50 & 0.30 & DP\\
0.01 & 1 & $-2\times 10^4$ & - & - & runs away\\
0 & 1 & $-10^4$ & 4.25 & 0.29 & weakly DP\\
0 & 1 & $-2\times 10^4$ & - & - & runs away\\
\hline
\end{tabular}
\end{flushleft}
\end{table}

In the present paper we have concentrated on the dynamics of the
current helicity (which is proportional to the function $\chi^c$)
and its comparison with the observations. At the present state of
observations we can compare latitudinal distributions of $\chi^c$
averaged over the activity cycle or the temporal behaviour of
$\chi^c$ averaged over a hemisphere. Such comparisons are
presented in Figs. 1--3 for $D=-10^3$, $\sigma =1$, $T=3$,
$(2R/l)^2=300$. In Figs. 1 and 3 the parameter  $\kappa = 0.1$. We
restrict ourselves to discussing dynamo models with singly
periodic behaviour (although there are hints of a double
periodicity in the sunspot record). Two types of behaviour are
demonstrated. Provided that $C$ is negative (magnetic helicity
inflow, see Fig. 2b) or small and positive ($C<0.1$, moderate
magnetic helicity outflow, see Fig. 2a) the value of $\chi^c$ is
always negative in the Northern hemisphere, in accordance with
naive theoretical expectations as well as the available
observations. If $C$ is large and positive (strong magnetic
helicity outflow), we obtain a cycle during which $\chi^c$ changes
sign. Both types of behaviour are shown in Fig.~1, for $C=0.01$
(solid curve) and $C=0.1$ (broken). Comparing observational and
theoretical results we have to fit the numerical data for $\chi^c$
with its observational equivalent $H_c$. We normalize the data to
make them equal at $\Theta = 28^\circ$ ($\langle \chi^c
\rangle_{\Theta = 28^\circ} = 23.5 \langle H_c \rangle$). Because
of this fitting procedure, the agreement between the observational
and numerical data is much better in the Northern hemisphere than
in the Southern.

\begin{figure}
\centering
\includegraphics[width=9cm]{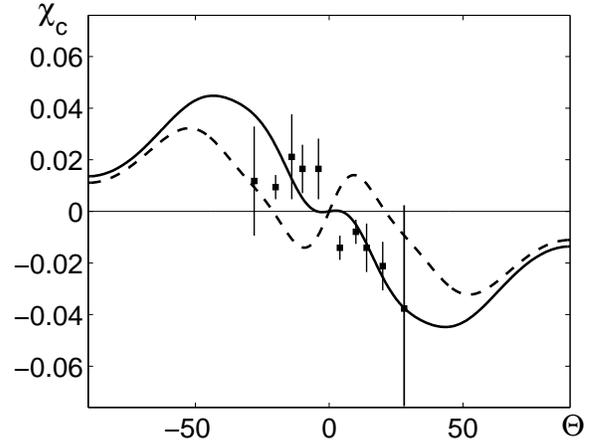}
\caption{\label{Fig1} The latitude dependence of the time-averaged
function $\langle \chi^c \rangle$ for various parameter sets (see
description in the text). The observed values of the time-averaged
$\langle H_c \rangle$ are shown by filled squares, the error-bars
are shown by vertical lines. A fitting factor of $23.5$, (i.e.,
$\langle \chi^c \rangle_{\Theta=28^\circ} = 23.5 \langle H_c
\rangle )$ has been used -- see text.}
\end{figure}

We illustrate the importance of the fitting procedure for various
parameter sets in Fig.~2 where we give the results for $C=0.01$ in
panel (a) and for $C=-0.1$ in panel (b). The results for $\kappa =
0.1$ are shown by broken curves and those for $\kappa=1$ by solid.
Both cases represent a model which does not exhibit helicity
reversals. We appreciate that the agreement between numerical and
observational data in the Northern hemisphere can be partially
attributed to the fitting procedure; however the agreement
obtained looks quite impressive for the primitive models
considered. A disagreement between the model and observations at
$\Theta = 4^\circ$ can be explained, for example, as a result of a
non-perfect North-South symmetry in the observed cycle. We
conclude from Fig.~2 that the minimal value of $\langle \chi^c
\rangle$ decreases with $\kappa$; experience from numerical
simulations show that it decreases with $\sigma$ as well. Note
however that the contribution of $\chi^c$ to the magnetic part of
the $\alpha$-effect is determined by $\sigma \langle \chi^c
\rangle$ and this value is more or less stable.

\begin{figure}
\centering
\includegraphics[width=8cm]{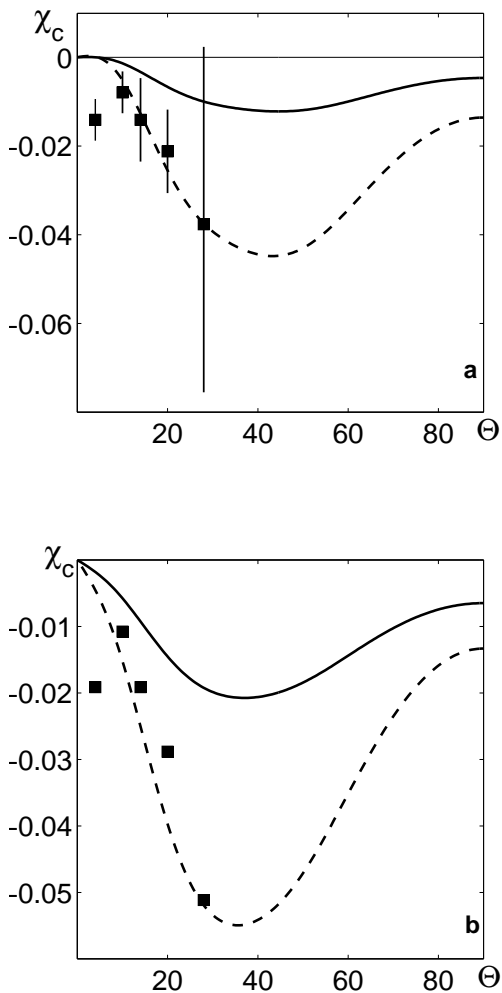}
\caption{\label{Fig2} The latitude dependence of the time-averaged
function $\langle \chi^c \rangle$ for various parameter sets (see
description in the text). The observed values of the time-averaged
$\langle H_c \rangle$ are shown by filled squares, the error-bars
are shown by vertical lines. We have used a fitting factor in (a)
of $23.5$, (i.e., $\langle \chi^c \rangle_{\Theta=28^\circ} = 23.5
\langle H_c \rangle )$ and in panel (b) it is $32$.}

\end{figure}

Moving to the comparison of the temporal helicity evolution
presented in Fig.~3, we note that the fitting procedures now have
to be more complicated. The first two measured points (1988-89)
and (1990-91) with approximately equal values of the
latitude-averaged $\langle H_c \rangle$ are chosen to be located
symmetrically about the minimum of the function $\langle \chi^c
\rangle$ for $\Theta > 0$. The distance between the observational
points is $ (2/11) (T_c/2)$, where $T_c$ is the period of
oscillations of the toroidal magnetic field. The fitting factor is
$44.3$, determined by the condition $\langle \chi^c
\rangle_{t=3.527 t_d} = 44.3 \langle H_c \rangle_{t=1988}$, where
$t_d$ is the turbulent diffusion time at the bottom of the
convective zone.

\begin{figure}
\centering
\includegraphics[width=9cm]{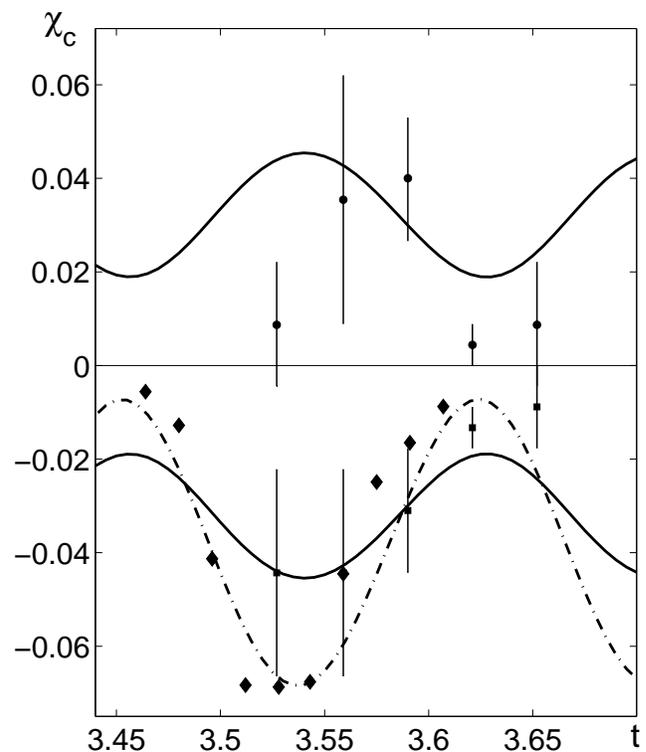}
\caption{\label{Fig3} The time dependence of the latitude-averaged
function $\langle \chi^c \rangle$ for $D=-10^3$, $\sigma=1$,
$C=0.01$ and $\kappa=0.1$. The observed values of the
latitude-averaged function $\langle H_c \rangle$ are shown by
filled squares (for $\Theta > 0$ -- Northern hemisphere, lower
panel) and filled circles (for $\Theta < 0$ -- Southern
hemisphere, upper panel), the error-bars for $\langle H_c \rangle$
are shown by vertical lines. The dashed-dotted line indicates the time
dependence of the latitude-averaged function $ - 6 \langle B^2
\rangle$. The filled diamonds in the lower panel give the scaled
averaged group sunspot numbers, $R_g$ -- see text for further
details.}
\end{figure}

The point is that we can base the fitting procedure on a
comparison between the simulated value of the toroidal magnetic
field and some tracer of cyclic solar activity, say, the averaged
group sunspot number $R_g$ (Hoyt et al. 1994)\footnote{The data for $R_g$ were
taken from the URL ftp.ngdc.noaa.gov/STP/}. This fitting is also shown
in Fig.~3 as follows. The dashed-dotted line shows the time
dependence of the latitude-averaged function $ - 6 \langle B^2
\rangle$ for the same parameters. The fitting factor $6$ is
determined by the condition $ 6 \langle B^2 \rangle_{t=3.527 t_d}
= 0.07 .$ The values $ - R_g / 2160 $ are shown in Fig.~3 by the
filled diamonds. The fitting factor $1/2160$ is determined by the
condition: $ R_g(t=1988) / 2160 = 0.07 .$
(Our slightly awkward looking choice of scale for some of the
quantities plotted in Fig.~3 arises because we choose to fit our
model in the Northern hemisphere where   $\langle \chi^c \rangle$ and
$\langle H_c \rangle$ are negative.)

It can be seen from Fig.~3 that the maxima of the magnetic energy
$\langle B^2 \rangle$ nearly coincide with the minima of the
function $\chi^c$. A similar behaviour was observed also for the
quantity $R_g$ (a tracer of cyclic solar activity) and $H_c$ (the
tracer of the current helicity), cf. Bao \& Zhang (1998).

\section{Discussion and conclusions}
We have presented above a simple model of the nonlinear solar
cycle with a dynamo saturation mechanism  based on magnetic
helicity conservation. The model is obviously oversimplified,
however it reproduces some features of real stellar cycles as far
as they are known from, say, the Wilson sample data (Baliunas et
al. 1995). In particular, we obtain both singly periodic and
doubly periodic cycles. As expected from qualitative arguments as
well from observational data (see e.g. Bruevich et al. 2001) the
transition from singly periodic to doubly periodic cycles is
associated with a general trend from cycles with smaller amplitude
to those with larger amplitude.

Our model of dynamo saturation gives stable oscillations for a
limited region of parameter space. If the dynamo generation
becomes stronger  the numerical solution runs away. On one hand,
we have not included in the model all kinds of nonlinear dynamo
saturation; e.g. buoyancy and spot formation obviously lead to
some losses of toroidal magnetic field and thus contribute to
dynamo saturation. In addition, increasing the dynamo number
can reduce the radial spatial scale of the toroidal magnetic field
and effectively enlarge the parameter $\mu$ which determines the
toroidal magnetic field dissipation. If $\mu$ is independent of
$D$, we thus  artificially overestimate the generation effect for
larger dynamo numbers. On the other hand, experience from dynamo
simulations as well as observational data suggests that cyclic
behaviour is typical for moderate dynamo action only and chaotic
behaviour occurs when the dynamo action is stronger. It is more
than natural to expect that a parameter range with chaotic
temporal behaviour will also exist in our model.

We have concentrated our attention here on current helicity data,
so a deeper comparison of the cycle parameters with observational
data is beyond the scope of this paper. Also, we should keep
in mind that we compare the simulated current helicity $ \propto
\chi^c$ with just one part of the surface current helicity
$\langle b_z(\vec{\nabla}\times \vec{b})_z \rangle$.

The properties of current helicity simulated with our model have
been compared with the available observational data. The results
of this comparison look quite promising in spite of the quite
limited extent of the observational data, as well as the crude
nature of the model. We have been able to choose a set of
governing parameters which give helicity properties comparable
with the available phenomenology. From another viewpoint, the
observational data are rich enough to indicate a disagreement
between the available observations and the predictions of the
model with other parameter sets.

The parameters which gives an agreement between simulations and
observations are quite plausible. However we feel that it is too
early to insist that this agreement is more than a coincidence.
Really, we base our comparison on the 10 year observational data
of one scientific team. We stress that an extension of the
observational programme to cover several cycles as well the
inclusion of data obtained by other scientific teams and from
other tracers would be very important. In particular it would be
valuable to include the cross-helicity data into the analysis.

Note, that although the available observations cover just the
period of 10 years, these data extend over parts of two different
solar cycles, namely the 22nd (1988-1995) and the 23rd
(1996-1997). The available data suggest that the shape of the
current helicity distribution is the same for both cycles, and
further, more recent, current helicity studies (Bao et al. 2000;
2002) support this interpretation. In particular, the hemispheric
rule is obeyed from cycle to cycle (e.g. Pevtsov et al. 2001).

We stress that the scenario described in the present paper does
not include all possible types of nonlinear processes which can
occur at the nonlinear stage of the dynamo (see, e.g., Brandenburg \&
Subramanian 2000; Brandenburg \& Dobler 2001), but rather is
restricted to a minimal number of processes involved in
magnetic helicity conservation. In the spirit of the basic ideas
about the nonlinear saturation of solar dynamos, the analysis
presented here has been restricted to the evolution of $\alpha$ only,
while detailed simulations (e.g. Blackman \& Brandenburg 2002;
Brandenburg \& Sokoloff 2002) also demonstrate a quenching of the
turbulent magnetic diffusivity. A quantitative model for a
nonlinear quenching of turbulent magnetic diffusivity has been
recently suggested by Rogachevskii \& Kleeorin (2001) and used in
a galactic dynamo model by Kleeorin et. al. (2003).

We appreciate that alternative interpretations for the current
helicity observations could be suggested, e.g. helicity could in
principle be produced during active region formation rather than
being connected with the magnetic field properties in the region
of intensive dynamo action. A development of such an alternative
explanation to the point where it could be confronted with the
observational data looks highly desirable.

We conclude with the expression of a guarded but real optimism
that magnetic helicity observations can result in the foreseeable
future in a new level of understanding in dynamo theory, which
will base the accepted $\alpha$-profiles not only on
order-of-magnitude arguments and numerical simulations, but also
on the observational data. It would be very important to support
this progress in solar dynamo theory by similar progress in other
areas of dynamo theory, in particular for galactic dynamos. We
mention in this respect a recent suggestion of En{\ss}lin \& Vogt
(2003) concerning the possibility of observing the magnetic
helicity of galactic magnetic fields.

\begin{acknowledgements}
Financial support from NATO under grant PST.CLG 974737, RFBR under
grants 03-02-16384, 01-02-17693, the RFBR-NSFC grant
02-02-39027 and the INTAS Program Foundation under grant 99-348 is
acknowledged. DS and KK are grateful for support from the Chinese
Academy of Sciences and NSFC towards their visits to Beijing.
\end{acknowledgements}

\appendix

\section{The nonlinear system of dynamo equations}
The system of nonlinear equations in nondimensional form for an
axisymmetric mean magnetic field $ \vec{B} = B \vec{e}_{\phi} +
\vec{\nabla} {\bf \times} (A \vec{e}_{\phi}) $ reads

\begin{eqnarray}
&&{\partial B \over \partial t} = D \, \hat \Omega A + \Delta_s B
\;,
\label{R1}\\
&&{\partial A \over \partial t} = \alpha B + \Delta_s A \;,
\label{R2}\\
&&{\partial \chi^c \over \partial t} + {\chi^c \over T} =
\biggl({2 R \over l} \biggr)^2 \biggl(\sin^2 \theta ({\vec
\nabla}_s A) \, ({\vec \nabla}_s B) - \, B \Delta_s A
\nonumber\\
&&- \alpha B^2 + {C \over R} \, \vec{\nabla} \cdot (l^2
\Lambda_\rho^{-1} \chi^v \, \phi_v \,B^2 \, \vec{e}_r) \biggr) +
\kappa \Delta \chi^c \;, \label{R3}
\end{eqnarray}
where $\alpha = \chi^v \phi_{v} + \chi^c \phi_{m} / \rho(r) ,$ $\,
\Delta_s = \sin^2 \theta \, {\vec \nabla}_s^2 \equiv \Delta - 1 /
(r^2 \, \sin^2 \theta) ,$ and

\begin{eqnarray*}
B_r &=& {R_\alpha \over r \, \sin \theta} {\partial \over \partial
\theta} (\sin \theta \, A) \;, \quad B_\theta = - {R_\alpha \over
r} {\partial \over \partial r} (r \, A) \;,
\\
{\vec \nabla}_s A &=& {1 \over r \, \sin \theta} \biggl({\vec e}_r
{\partial \over \partial r} (r A) + {\vec e}_\theta \, {1 \over
\sin \theta} {\partial \over \partial \theta} (\sin \theta \, A)
\biggr) \;,
\\
\hat \Omega A &=& {1 \over r} {\partial (\Omega , A r \sin \theta)
\over \partial (r , \theta)} \; .
\end{eqnarray*}
Now we consider dynamo action in a thin convective shell, average
$A$, $\, B$ and $\alpha$ over the depth of the shell and
consider these quantities as functions of colatitude $\theta$
only. Then we neglect the convection shell curvature and replace
it by a flat slab. This implies that $\Delta_s = \Delta =
\partial^2 / \partial \theta^2 - \mu^2 ,$ $\, \sin \theta
{\vec \nabla}_s = {\vec e}_\theta (\partial / \partial \theta) +
\mu {\vec e}_r $ and $\, \hat \Omega A = g (\partial A / \partial
\theta) .$ We also redefine $ C \, \mu \, l^2 / \Lambda_\rho \, R$
as $C$. This yields Eqs.~(\ref{eqB}), (\ref{eqA}) and~(\ref{E9}).

\end{document}